\documentclass[aps,pra,twocolumn,superscriptaddress,nofootinbib]{revtex4-2}
\usepackage{graphics}
\usepackage{subfigure}
\usepackage{amsmath}
\usepackage{physics}
\usepackage{graphicx}
\usepackage{color}
\usepackage{amsfonts}
\usepackage{amssymb}
\usepackage{float}
\usepackage{hyperref}

\DeclareMathOperator*{\argmax}{arg\,max}

\begin{document}

\title{Bounding the benefit of adaptivity in quantum metrology using the relative fidelity}

\author{Jason L. Pereira} \email{jason.pereira@york.ac.uk}
\affiliation{Department of Computer Science, University of York, York YO10 5GH, UK}
\affiliation{Department of Physics and Astronomy, University of Florence,
via G. Sansone 1, I-50019 Sesto Fiorentino (FI), Italy}

\author{Leonardo Banchi}
\affiliation{Department of Physics and Astronomy, University of Florence,
via G. Sansone 1, I-50019 Sesto Fiorentino (FI), Italy}
\affiliation{ INFN Sezione di Firenze, via G. Sansone 1, I-50019, Sesto Fiorentino (FI), Italy }

\author{Stefano Pirandola}
\affiliation{Department of Computer Science, University of York, York YO10 5GH, UK}

\date{\today}

\begin{abstract}
	Protocols for discriminating between a pair of channels or for estimating a channel parameter can often be aided by adaptivity or by entanglement between the probe states. This can make it difficult to bound the best possible performance for such protocols. In this paper, we introduce a quantity that we call the relative fidelity of a given pair of channels and a pair of input states to those channels. Constraining the allowed input states to all pairs of states whose fidelity is greater than some minimum ``input fidelity" and minimising this quantity over the valid pairs of states, we get the minimum relative fidelity for that input fidelity constraint. We are then able to lower bound the fidelity between the possible output states of any protocol acting on one of two possible channels in terms of the minimum relative fidelity. This allows us to bound the performance of the most general, adaptive discrimination and parameter estimation protocols. By finding a continuity bound for the relative fidelity, we also provide a simple confirmation that the quantum Fisher information (QFI) of the output of an $N$-use protocol is no more than $N^2$ times the one-shot QFI.
\end{abstract}

\maketitle

\section{Introduction}

Many physical processes can be modelled as quantum channels, so determining the identity of an unknown channel and estimating a parameter encoded in that channel are important tasks in the field of quantum information. The quantum fidelity~\cite{nielsen_quantum_2011,fidelity2,uhlmann_transition_1976} between the possible output states of channel discrimination or parameter estimation protocols give an indication of the distinguishability of those output states. By bounding the quantum fidelity between the possible output states of a protocol, we can bound the performance of both quantum channel discrimination~\cite{helstrom_quantum_1976,wang_unambiguous_2006,chefles_quantum_2000,barnett_quantum_2009,sun_optimum_2001,bergou_optimal_2012,pirandola_advances_2018,pirandola_fundamental_2019,zhuang_ultimate_2020,zhuang_entanglement-enhanced_2020,banchi2020quantum,harney2020ultimate,pereira2020idler} and quantum metrology~\cite{braunstein_statistical_1994,acin_optimal_2001,paris_quantum_2009,giovannetti_advances_2011,RevModPhys.90.035005,JasReview,braun_quantum-enhanced_2018,pirandola_advances_2018}.

To achieve the minimum possible output fidelity, we must allow the protocols to be adaptive. This means that the output from a previous use of the unknown channel can affect the input to a subsequent channel use. In an example with two discrete-variable channels, it was shown that adaptive schemes can beat non-adaptive schemes for channel discrimination~\cite{harrow_adaptive_2010}.

Some channel pairs can be perfectly discriminated after finite uses (such as unitaries or the channels in Ref.~\cite{harrow_adaptive_2010}), whilst others can never be perfectly discriminated after a finite number of channel uses, even if adaptivity or entangled probes can reduce the error probability (such as classical channels~\cite{sacchi_entanglement_2005,harrow_adaptive_2010,hayashi_discrimination_2009}). Ref.~\cite{duan_perfect_2009} gives necessary and sufficient conditions for distinguishing between the two cases. Some work has been done to bound the asymptotic benefit of adaptivity~\cite{wilde_amortized_2020}, but less is known about adaptivity for finite channel uses.

Here, we present a new measure on a pair of channels and a pair of input states to those channels called the relative fidelity. If we constrain the allowed input states to have a fidelity greater than or equal to some minimum ``input fidelity" and then minimise the relative fidelity over the allowed states (for a fixed pair of channels), we get the minimum relative fidelity for that channel pair and that input fidelity. We can use this quantity to formulate lower bounds on the minimum fidelity between output states for any adaptive protocol. If the minimum relative fidelity is constant (i.e. has no dependence on the input fidelity), the optimal protocol not only does not require adaptivity but also does not require entanglement between the input states for each channel use. Otherwise, we can use the minimum relative fidelity to formulate bounds on the performance of any protocol that hold for finite uses as well as asymptotically. By finding continuity bounds on the minimum relative fidelity, we provide a simple confirmation that the maximum quantum Fisher information (QFI) for an $N$-use protocol is no more than $N^2$ times the maximum one-shot QFI~\cite{pirandola_fundamental_2019,zhou_asymptotic_2020,katariya_geometric_2021}.

\section{Bounding the minimum fidelity between the outputs of a protocol}

Suppose we have a black box containing a channel, $\mathcal{C}$, drawn from a set of two possible channels, $\{\mathcal{C}_1,\mathcal{C}_2\}$, both of which have input dimension $d$. Our task is to achieve the minimum fidelity between the two possible outputs of a fixed protocol that involves $N$ uses of $\mathcal{C}$. This protocol can be adaptive, meaning that the input for a channel use can depend on the output from every previous channel use. Such protocols can be represented as quantum combs~\cite{chiribella_quantum_2008,pirandola_advances_2018} and are the most general strategies allowed by quantum mechanics.

Lower bounding the minimum fidelity between protocol outputs allows us to upper bound the distinguishability of the channels $\mathcal{C}_1$ and $\mathcal{C}_2$. The minimum probability of error in discriminating between a pair of states is bounded by the Fuchs-van der Graaf inequality~\cite{fuchs_cryptographic_2006}:
\begin{equation}
    p_{\mathrm{err}}\geq\frac{1-\sqrt{1-F^2}}{2},\label{eq: fuchs-van der graaf}
\end{equation}
where $F$ is the fidelity between the states. Consequently, the minimum fidelity between the output states of any $N$-use protocol lower bounds the error probability for any such protocol that discriminates between the channels.

Alternatively, suppose we have a family of channels parametrised by a variable $\theta$, $\mathcal{C}_{\theta}$, and can find an analytical expression for the minimum fidelity between the outputs of an $N$-use protocol acting on channels $\mathcal{C}_{\theta}$ and $\mathcal{C}_{\theta+\Delta_\theta}$, in terms of $\theta$. We can use this expression to upper bound the achievable QFI. The QFI is a crucial quantity in quantum metrology, because it appears in the quantum Cram\'{e}r–Rao bound (QCRB), which bounds the variance of parameter estimation~\cite{braunstein_statistical_1994}. The QCRB states that
\begin{equation}
    \mathrm{var}(\Tilde{\theta})\geq\frac{1}{\mathrm{QFI}_N(\theta)},
\end{equation}
where $\Tilde{\theta}$ is an unbiased estimator of $\theta$, so by upper bounding the QFI of any possible protocol output state, we lower bound the variance of an unbiased estimator of $\theta$.

\begin{figure}[tb]
\vspace{+0.1cm}
\centering
\includegraphics[width=0.75\linewidth]{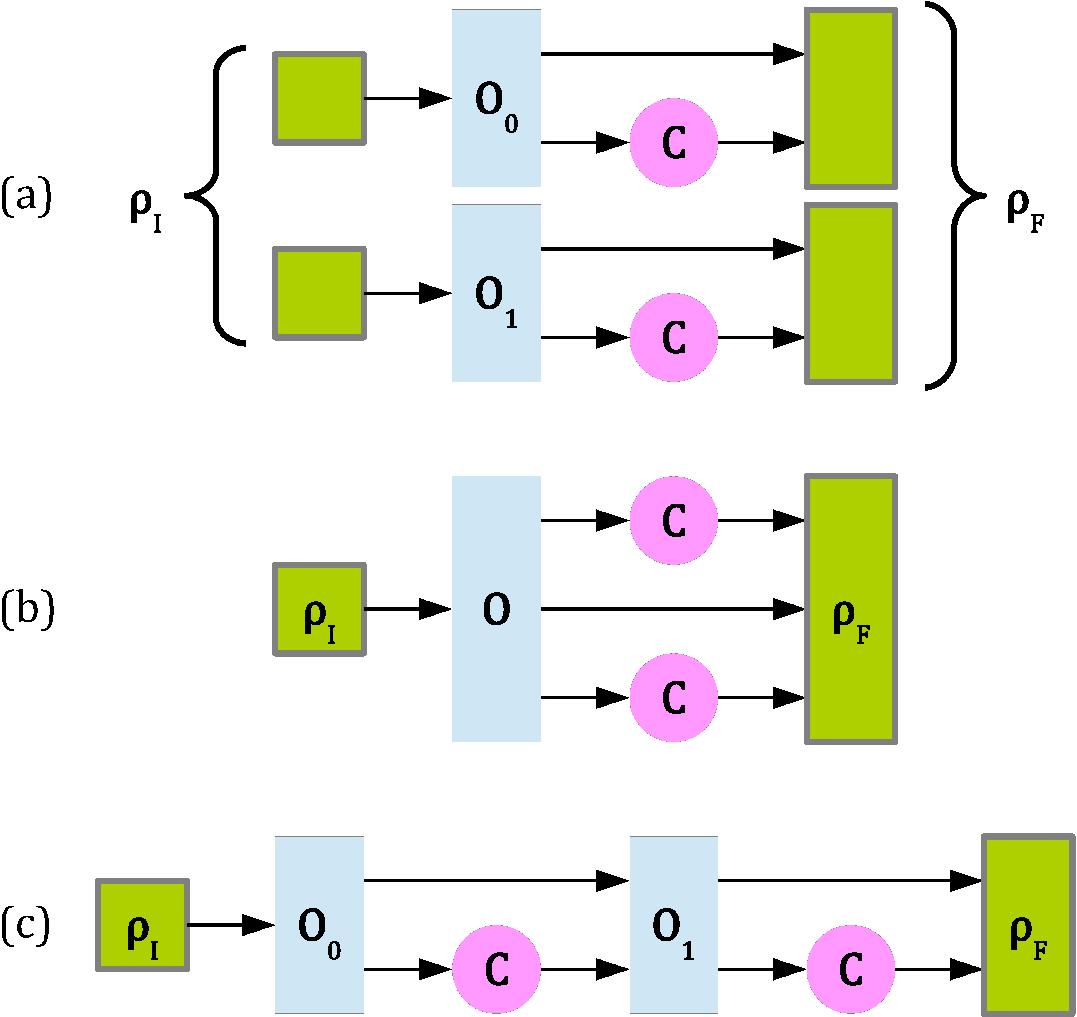}\caption{Setups for different types of protocols involving two channel uses. In all cases, there is a channel-independent initial state, $\rho_I$, which undergoes state preparation by means of some unitary and is subject to a total of two channel uses (labelled $C$), resulting in a final state, $\rho_F$. In (a), state preparation is carried out in such a way that $\rho_F$ takes tensor product form (there is no entanglement between the probe states for each channel use). $O_0$ prepares the first probe state and $O_1$ prepares the second. In (b), the initial state preparation, $O$, can result in entanglement between probe states, but there is no feedback between the output of one channel use and the input of the other. (b) defines the set of parallel protocols. In (c), we have a fully adaptive strategy, with $O_0$ carrying out the initial state preparation and $O_1$ describing any arbitrary processing on the state prior to the second channel use. We note that all operations can be represented as unitary operations on larger ancillary systems, via the principle of deferred measurement. The protocols in (a) are a subset of those in (b), which are themselves a subset of those in (c).}
\label{fig: setup}
\end{figure}

Fig.~\ref{fig: setup}  illustrates, for two channel uses, the different types of protocols that we consider. In the simplest case, we have independently prepared probe states (with no shared entanglement), each of which passes through the channel, resulting in a final state that takes tensor product form. Each probe can have idler systems (systems which do not pass through the channel, but may be entangled to states which do). The second case is similar, but entanglement between the probe states is allowed. The final case is fully adaptive. Note that all of the operations, $O_i$, can be regarded as unitaries on a larger Hilbert space, with the extra ancillary systems traced over at the end. We are interested in classifying for which channel discrimination problems there are protocols in (c) that are more powerful than all protocols in (a).

\subsection{Defining relative fidelity}

The fidelity between a pair of states is defined by
\begin{equation}
    F(\rho_1,\rho_2)=\mathrm{Tr}\left[\sqrt{\sqrt{\rho_1}\rho_2\sqrt{\rho_1}}\right].
\end{equation}
We define the output fidelity, $F_{\mathrm{out}}$, between $\mathcal{C}_1$ and $\mathcal{C}_2$, for a given pair of input states, $\sigma_1$ and $\sigma_2$, as the fidelity between the channel outputs. In other words
\begin{equation}
    F_{\mathrm{out}}(\sigma_1,\sigma_2)=F(\mathcal{I}\otimes\mathcal{C}_1[\sigma_1],\mathcal{I}\otimes\mathcal{C}_2[\sigma_2]),
\end{equation}
where the identity operator acts on the idler modes. Let $F_{\mathrm{con}}$ be the minimum output fidelity for constant input
\begin{equation}
    F_{\mathrm{con}}=\inf_{\sigma\in \mathcal{D}(d^2)} F_{\mathrm{out}}(\sigma,\sigma),
\end{equation}
with $\mathcal{D}(d^2)$ being the space of density operators of dimension $d^2$. Note that $F_{\rm con}$ can be efficiently computed via semidefinite programming~\cite{yuan2017quantum}.

Let us consider how the fidelity between output states evolves at each stage of the protocol. We define $F_{N}$ as the fidelity between the possible output states after $N$ channel uses, considering the most general case shown in Fig.~\ref{fig: setup}(c). Similarly, $F_{i}$, with $i<N$, is the fidelity between the possible output states that would be obtained if the protocol were terminated prematurely immediately prior to the $(i+1)$-th channel use. Fidelity is non-decreasing over the (adaptive) trace-preserving operations $O_i$ between each channel use, while it can (only) be reduced by a use of the channel $\mathcal{C}$, because this is different for each possible output. It is immediate from the multiplicativity of fidelity over tensor products that we can always reduce the output fidelity by a multiplicative factor of at least $F_{\mathrm{con}}$ at each channel use, i.e. we can always choose some input to the $i$-th channel use such that
\begin{equation}
    F_{i}\leq F_{\mathrm{con}}F_{i-1}.
\end{equation}
We can often do better than this, even for parallel strategies (strategies that do not involve adaptivity). For certain pairs of channels, the fidelity between output states can be multiplied by a factor smaller than $F_{\mathrm{con}}$ each time the channel is applied to one part of the probe state. An example would be discriminating between unitaries using some initially entangled, multipartite probe state, such as the GHZ state (this would be an example of a protocol in (b), from Fig.~\ref{fig: setup}). Such protocols can perfectly discriminate between two unitaries after finite channel uses~\cite{acin_statistical_2001,dariano_using_2001}.

A natural question is ``how much can the fidelity between output states be reduced by in a single channel use?" Equivalently, we can ask ``what is the minimum achievable value of $\frac{F_{i-1}}{F_{i}}$?" To address this question, we introduce the ``relative fidelity", $F_R$, which we define as
\begin{equation}
    F_R(\sigma_1,\sigma_2)=\frac{F_{\mathrm{out}}(\sigma_1,\sigma_2)}{F(\sigma_1,\sigma_2)}.
\end{equation}
The relative fidelity is the ratio between the output fidelity and the input fidelity for a pair of states, $\sigma_1$ and $\sigma_2$. Note that this can be either greater than or less than (or equal to) one, depending on the choice of states. In order for $F_R$ to be well-defined for all inputs, we define
\begin{equation}
    F_R(\sigma_1,\sigma_2)=\lim_{\delta\to0}F_R(\sigma_1,(1-\delta)\sigma_2+\delta\sigma_1)
\end{equation}
for orthogonal $\sigma_1$ and $\sigma_2$ (since otherwise we would have both the numerator and the denominator equal to zero).

We now define the quantity
\begin{equation}
    F_{R,\mathrm{min}}(f)=\inf_{\{\sigma_1,\sigma_2\}\in \{\mathcal{D}:F(\sigma_1,\sigma_2)\geq f\}} F_R(\sigma_1,\sigma_2),
\end{equation}
where $\{\mathcal{D}:F(\sigma_1,\sigma_2)\geq f\}$ is the set of all pairs of density matrices with a fidelity $\geq f$. $F_{R,\mathrm{min}}(f)$ gives the maximum amount that the output fidelity of a protocol can be reduced by with a single channel use (the minimum factor that it can be multiplied by), so long as the fidelity before that channel use was $\geq f$. In the case in which $f=0$, this becomes the fidelity divergence from Ref.~\cite{chiribella_quantum_2019}, and is also equivalent to the amortised channel divergence from Ref.~\cite{wilde_amortized_2020} (by choosing the generalised divergence to be the sandwiched R\'{e}nyi entropy with $\alpha=\frac{1}{2}$, in Eq.~(52), and exponentiating). In both cases, the optimisation has no input fidelity constraint, so these quantities define the asymptotics of the output fidelity. For more information on the differences between $F_{R,\mathrm{min}}(f)$ and quantities used in previous works, see Appendix~\ref{appendix differences}.

If $F_{R,\mathrm{min}}(f)<F_{\mathrm{con}}$, this does not necessarily mean adaptivity is required for the optimal protocol. $F_{R,\mathrm{min}}(f)$ may not be achievable by any protocol. For instance, the channels could output mixed states, but the inputs giving $F_{R,\mathrm{min}}(f)$ could be pure. Alternatively, a protocol may be non-adaptive but have entanglement between the probe states (as for the GHZ probe). This is still non-adaptive, despite the shared input state, because the output of one channel use never affects the input to another. In fact, for unitary channel discrimination, it is known that a non-adaptive strategy with entanglement between the probe states can be used instead of an adaptive strategy in order to achieve optimal discrimination~\cite{dariano_using_2001}.

The opposite is not true. If $F_{R,\mathrm{min}}(F_{\mathrm{con}}^{N-1})=F_{\mathrm{con}}$, then no $N$-use protocol can benefit from adaptivity or entanglement between the probe states (note that $F_{R,\mathrm{min}}(F_{\mathrm{con}}^{N-1})$ is never more than $F_{\mathrm{con}}$, since we can always choose $\sigma_1$ and $\sigma_2$ to both be $\sigma_{\mathrm{min}}$, the state that achieves $F_{\mathrm{con}}$). If $F_{R,\mathrm{min}}(0)=F_{\mathrm{con}}$, adaptivity (and entanglement between probe states) is never required for the optimal protocol, and we can write the following inequality for any $N$-use protocol
\begin{equation}
    F_N \geq F_{\mathrm{con}}^{N}.
\end{equation}
Otherwise, we can write the lower bound:
\begin{equation}
    F_{N}\geq F_{N-1}F_{R,\mathrm{min}}(F_{N-1}).\label{eq: adaptive LB}
\end{equation}
Numerically, this can be calculated recursively, starting from a single channel use. The lower bound on $F_N$ can then be substituted into Eq.~(\ref{eq: fuchs-van der graaf}) to lower bound the error probability for discrimination between a pair of channels. If the channel pair is perfectly distinguishable, we can lower bound the minimum number of uses for a protocol to do so by finding the smallest value of $N$ such that we can have $F_{R,\mathrm{min}}(F_{N-1})=0$.

One question, when calculating $F_{R,\mathrm{min}}(f)$, is: what is the maximum dimension of $\sigma_1$ and $\sigma_2$? In the case of a constant input, a pure state is optimal, due to the superadditivity of fidelity, and the maximum required dimension of the input is $d^2$. For $f<1$, it is not immediately obvious, but we can again show that $F_{R,\mathrm{min}}(f)$ can always be achieved by a pair of pure input states with dimension $d^2$. The proof of this is given in Appendix~\ref{appendix property}.

A key property that the relative fidelity lacks (but that fidelity has) that may make the task of finding the minimum relative fidelity more difficult is concavity. See the supplementary MATLAB files for a counterexample based on a pair of amplitude damping channels~\cite{SUPP}.

\section{Continuity properties}

Suppose we know the minimum relative fidelity for some input fidelity, $f'$, and wish to bound its value for some other input fidelity, $f$. We can do so using the continuity bound
\begin{equation}
    F_{R,\mathrm{min}}(f)\geq \frac{1 - \left(\sqrt{1-f' F_{R,\mathrm{min}}(f')} +\sqrt{f'-f}\right)^2}{f},\label{eq: min_rel_fid continuity}
\end{equation}
where $f<f'$. This recursively defines $F_{R,\mathrm{min}}(f)$ in terms of $F_{R,\mathrm{min}}(f')$. See Appendix~\ref{appendix continuity} for the proof of this bound. A scheme for numerically finding the minimum relative fidelity for any pair of channels, based on an output fidelity continuity bound, is given in Appendix~\ref{appendix algorithm}.

Setting $f'=1$, in Eq.~(\ref{eq: min_rel_fid continuity}), we lower bound $F_{R,\mathrm{min}}(f)$ for any channel. We can write
\begin{equation}
    F_{\mathrm{con}}\geq F_{R,\mathrm{min}}(f)\geq \frac{1 - \left(\sqrt{1- F_{\mathrm{con}}}+\sqrt{1-f}\right)^2}{f},\label{eq: adaptivity bound}
\end{equation}
where $F_{\mathrm{con}}$ is, as before, the minimum output fidelity for constant channel inputs. Whilst Eq.~(\ref{eq: adaptivity bound}) is not (necessarily) tight, it gives an ultimate bound on the benefit of adaptivity for any channel. The lower bound corresponds to the best possible output fidelity scaling that any protocol can achieve, adaptive or otherwise.

We may instead choose to express this bound in terms of the minimum output fidelity for a given input fidelity:
\begin{equation}
    F_{\mathrm{out},\mathrm{min}}(f)\geq 1 - \left(\sqrt{1- F_{\mathrm{con}}}+\sqrt{1-f}\right)^2.\label{eq: min N-use output fidelity}
\end{equation}
This can also be expressed in terms of the minimum output fidelity of an $N$-use protocol as
\begin{equation}
    F_N\geq 1-N^2(1-F_{\mathrm{con}}).\label{eq: max fidelity scaling}
\end{equation}
This can be verified by substituting the right-hand side of Eq.~(\ref{eq: max fidelity scaling}) as $f$ in Eq.~(\ref{eq: min N-use output fidelity}).

The $N$-use QFI is given by~\cite{braunstein_statistical_1994,liu_fidelity_2014}
\begin{equation}
    \mathrm{QFI}_N(\theta)=\frac{8(1-F_N(\theta,\theta+d\theta))}{d\theta^2}.\label{eq: QFI}
\end{equation}
We can write
\begin{equation}
    1-F_{N}(\theta,\theta+d\theta)\leq N^2(1-F_{\mathrm{con}}(\theta,\theta+d\theta)),
\end{equation}
with equality if, for every $\theta$ and $\theta+d\theta$, we can achieve the $N$-use output fidelity given in Eq.~(\ref{eq: max fidelity scaling}) (with $F_{\mathrm{con}}$ depending on the choice of $\theta$). Using Eq.~(\ref{eq: QFI}), we write
\begin{equation}
    \mathrm{QFI}_N(\theta)\leq \frac{8N^2(1-F_{\mathrm{con}}(\theta,\theta+d\theta))}{d\theta^2}=N^2\mathrm{QFI}_1(\theta),
\end{equation}
where $\mathrm{QFI}_1$ is the one-shot QFI. Consequently, we have a simple confirmation (via the QCRB) that the variance scales with the inverse square of the number of channel uses~\cite{pirandola_fundamental_2019,zhou_asymptotic_2020,katariya_geometric_2021}. We have therefore recovered the Heisenberg scaling as the maximum possible scaling with the number of channel uses, as expected~\cite{giovannetti_advances_2011}.

\section{Scaling with input fidelity}

\begin{figure}[tb]
\vspace{+0.1cm}
\centering
\includegraphics[width=0.9\linewidth]{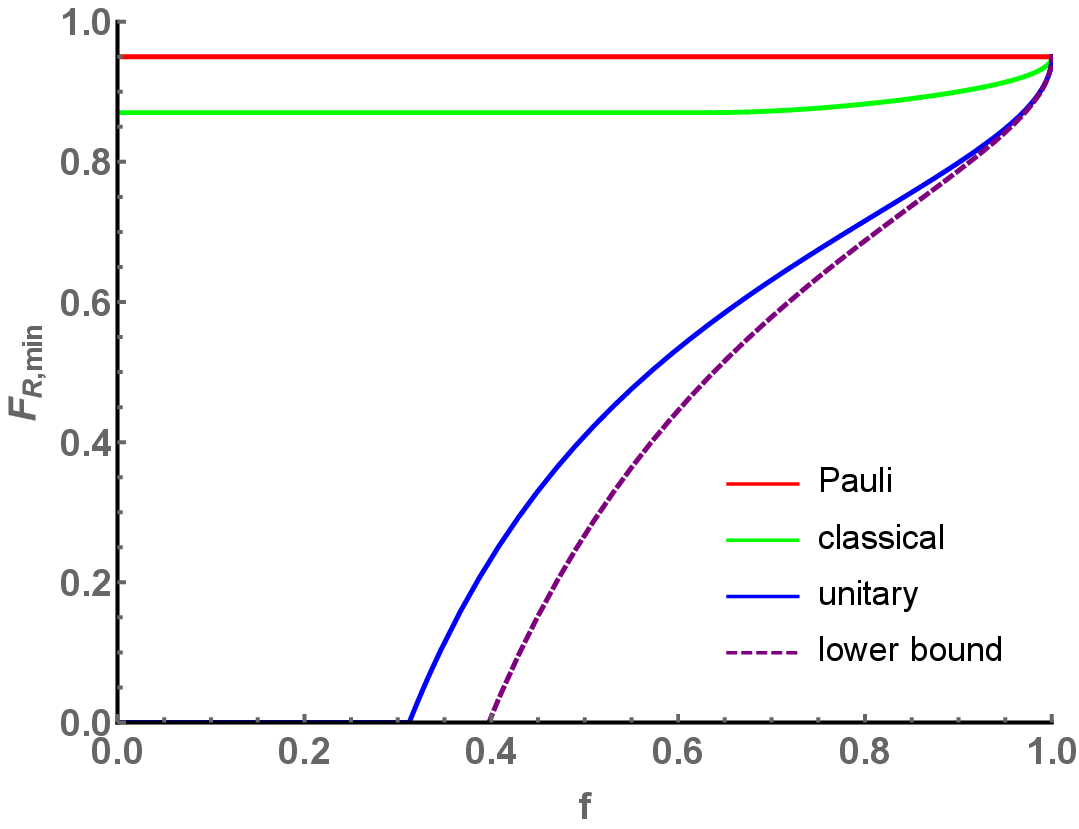}\caption{The different ways in which the minimum relative fidelity, $F_{R,\mathrm{min}}$, could scale with the input fidelity, $f$. The three continuous lines represent specific examples of channel pairs demonstrating different types of scaling, whilst the dashed line is the lower bound in Eq.~(\ref{eq: adaptivity bound}). All three examples have $F_{\mathrm{con}}=0.95$. The top line represents a pair of Pauli channels, for which neither adaptivity nor entanglement between the probes can reduce the minimum fidelity between output states. It corresponds to the upper bound in Eq.~(\ref{eq: adaptivity bound}). The middle line represents a pair of classical channels, for which adaptivity or entanglement between probes may reduce the output fidelity but perfect discrimination is still impossible. The bottom line represents a pair of unitary channels for which adaptivity or entanglement between probes can reduce the output fidelity and perfect discrimination is possible.}
\label{fig: scaling}
\end{figure}

There are three possible scenarios for the scaling of $F_{R,\mathrm{min}}$ with the input fidelity.

The first is where $F_{R,\mathrm{min}}$ is constant and equal to $F_{\mathrm{con}}$. This corresponds to channel discrimination problems for which neither adaptivity nor entanglement between probe states are required for optimal discrimination (and perfect discrimination is not possible). Discrimination between Pauli channels is a known example of such a problem~\cite{pirandola_fundamental_2017}.

Secondly, there are cases in which $F_{R,\mathrm{min}}$ decreases with the input fidelity and eventually goes to zero. In these situations, it may be the case that adaptivity or entanglement between probe states (or both) are required for optimal discrimination, and perfect discrimination may be possible. An example of such a situation is discrimination between two unitaries~\cite{acin_statistical_2001,dariano_using_2001}.

Finally, for some discrimination tasks, $F_{R,\mathrm{min}}$ may decrease from $F_{\mathrm{con}}$ but then tend to a non-zero value at an input fidelity of zero (so that $0<F_{R,\mathrm{min}}(0)<F_{\mathrm{con}}$). This implies that adaptivity or entanglement between probe states may be required for optimal discrimination, but that perfect discrimination is not possible. Entanglement-breaking (classical) channels are an example of this scenario~\cite{sacchi_entanglement_2005,hayashi_discrimination_2009,harrow_adaptive_2010}.

By investigating simple examples of each scenario and analytically finding the minimum relative fidelity, we can confirm that the minimum relative fidelity acts as expected in these situations. Fig.~\ref{fig: scaling} shows the relative fidelity scaling for each of the examples. These examples are discussed in more detail in Appendix~\ref{appendix examples}, and demonstrate how the relative fidelity can be used to bound how the channel discrimination error probability scales with the number of channel uses and identify situations in which adaptivity and entanglement between probes cannot help with channel discrimination.

\section{Conclusion}
In this paper, we introduced a quantity that we call the relative fidelity of a pair of channels, which is the ratio between the input and the output fidelity for a pair of states. By minimising this quantity, subject to a constraint on the input fidelity, we lower bound the fidelity between outputs of an $N$-use channel discrimination or parameter estimation protocol. If the minimum relative fidelity is non-zero for every input fidelity, the pair of channels can never be perfectly discriminated.

The minimum relative fidelity for a given pair of channels (acting on $d$-dimensional inputs) and a given input fidelity constraint can always be achieved by a pair of pure states, each with dimension no more than $d^2$. We found a minimum relative fidelity continuity bound, and so lower bounded the output fidelity for any adaptive protocol discriminating between any pair of channels. As a result, we demonstrated that the QFI for an N-use protocol is no more than $N^2$ times the maximum one-shot QFI~\cite{pirandola_fundamental_2019,zhou_asymptotic_2020,katariya_geometric_2021}, via a simple proof.

The minimum relative fidelity is a quantity that could prove useful for providing ultimate bounds on the performance of channel discrimination or parameter estimation protocols. Because finding it only requires a maximisation over two $d^2$-dimensional pure qudit states for a given input fidelity, it should be easier, in many cases, to bound the performance of a protocol in this way than to optimise over all possible adaptive protocols (as represented by the set of all $d^{2N}$-dimensional quantum combs). Although the result of a full optimisation would be more precise than a bound based on the minimum relative fidelity, minimisation over the set of quantum combs involves $\sim d^{4N}$ free parameters (since each comb can be represented by a $d^{2N}$ by $d^{2N}$ Choi matrix), meaning that such minimisations quickly become difficult as $N$ increases. On the other hand, $F_{R,\mathrm{min}}(0)$ may not provide a good bound until $N$ is large, so Eq.~(\ref{eq: adaptive LB}) could provide a significantly tighter bound than using the asymptotic value.

\smallskip
\textbf{Acknowledgments.}~This work was funded by the European Union's Horizon 2020 Research and Innovation Action under grant agreement No. 862644 (FET-OPEN project: Quantum readout techniques and technologies, QUARTET). L.B.~acknowledges support by the program ``Rita Levi  Montalcini'' for
young researchers.

\appendix

\section{Proof the minimum relative fidelity can be achieved by a pair of pure states with dimension $d^2$}\label{appendix property}

It is possible to prove that the minimum relative fidelity for any input fidelity is achieved by a pure state, by using an alternative definition of the fidelity. The fidelity between any two states is the maximum absolute value of the overlap between purifications. This can be written as
\begin{equation}
    F(\rho_1,\rho_2)=\max_{\left|\rho_1'\right>,\left|\rho_2'\right>} \left|\left<\rho_1'\middle|\rho_2'\right>\right|,
\end{equation}
where $\left|\rho_{1(2)}'\right>$ is a purification of $\rho_{1(2)}$. Note that, for pure states,
\begin{equation}
    F(\rho_1,\rho_2)=\left|\left<\rho_1'\middle|\rho_2'\right>\right|.
\end{equation}
We therefore know that, for any pair of states, there exists some purification that does not decrease (or increase) the fidelity between channel outputs. Starting from the input states $\sigma_1$ and $\sigma_2$, let us replace them with the purifications that maximise their overlap, $\left|\sigma_1'\right>$ and $\left|\sigma_2'\right>$. These new input states have the same input fidelity. Since tracing over modes can never decrease the fidelity between a pair of states, the new input states also achieve the same or a lower output fidelity. Thus, $\left|\sigma_1'\right>$ and $\left|\sigma_2'\right>$ obtain a relative fidelity that is lower than or equal to that obtained by $\sigma_1$ and $\sigma_2$. Consequently, pure input states obtain the minimum relative fidelity for  given input fidelity. Note that the same proof was used in Ref.~\cite{chiribella_quantum_2019} to show that $F_{R,\mathrm{min}}(0)$ can be achieved using pure input states.

Now suppose $\sigma_1$ is pure. Let us write
\begin{equation}
    \sigma_1 =(\sigma_1)_{SI},
\end{equation}
where $S$ labels the system that passes through the channel and $I$ labels the idler modes (with unbounded dimension). There exists some unitary $U$, acting only on the idler modes, such that we obtain
\begin{equation}
    U\sigma_1 U^{\dagger} =(\sigma_1')_{SI'}\otimes\left|0\right>\left<0\right|_{I''},
\end{equation}
where $I'$ labels a subset of the idler modes with dimension $d$ and $I''$ labels some other subset of the idler modes with unbounded dimension. If $U$ is applied to $\sigma_2$ as well, neither the fidelity of the input states nor the output fidelity will be affected, and so the relative fidelity will be unchanged. We call the obtained state $\sigma_2'$. We now write
\begin{equation}
    \sqrt{(\sigma_1')_{SI'}\otimes\left|0\right>\left<0\right|_{I''}}=\left(\sqrt{\sigma_1'}\right)_{SI'}\otimes\left|0\right>\left<0\right|_{I''}.
\end{equation}
Now we note that
\begin{equation}
    (\sigma_2'')_{SI'}=\frac{1}{\alpha}\left<0\right|_{I''}(\sigma_2')_{SI'I''}\left|0\right>_{I''},
\end{equation}
where $\alpha$ is a normalising factor, is a state on the $d^2$-dimensional system $SI'$. $\alpha$ is given by
\begin{equation}
    \alpha=\mathrm{Tr}\left[\left<0\right|_{I''}(\sigma_2')_{SI'I''}\left|0\right>_{I''}\right].
\end{equation}
If $\sigma_2$ is pure then $\sigma_2''$ will also be pure. Consequently, we can write
\begin{equation}
    \begin{split}
        F(\sigma_1,\sigma_2)&=\mathrm{Tr}\left[\sqrt{\sqrt{\sigma_1}\sigma_2\sqrt{\sigma_1}}\right]\\
        &=\mathrm{Tr}\left[\sqrt{\left(\sqrt{\sigma_1'}\otimes\left|0\right>\left<0\right|\right)\sigma_2'\left(\sqrt{\sigma_1'}\otimes\left|0\right>\left<0\right|\right)}\right]\\
        &=\mathrm{Tr}\left[\sqrt{\sqrt{\sigma_1'}\left<0\right|_{I''}(\sigma_2')_{SI'I''}\left|0\right>_{I''}\sqrt{\sigma_1'}}\right]\\
        &=\sqrt{\alpha}F(\sigma_1',\sigma_2'').
    \end{split}
\end{equation}
Then, since the channels are trace-preserving, we can also write
\begin{equation}
    F_{\mathrm{out}}(\sigma_1,\sigma_2)=\sqrt{\alpha}F_{\mathrm{out}}(\sigma_1',\sigma_2'').
\end{equation}
Therefore,
\begin{equation}
    F_R(\sigma_1,\sigma_2)=F_R(\sigma_1',\sigma_2''),
\end{equation}
and so the relative fidelity of the dimension unbounded states $\sigma_1$ and $\sigma_2$ is the same as that of some other pair of states, $\sigma_1'$ and $\sigma_2''$, which each have a dimension of $d^2$.

Combining the two results, we can see that the minimum relative fidelity, $F_{R,\mathrm{min}}(f)$, can always be obtained by a pair of pure input states with dimension $d^2$.

Although we do not guarantee that the states that achieve $F_{R,\mathrm{min}}(f)$ (which we will call $\left|\sigma_{1,\mathrm{min}}\right>$ and $\left|\sigma_{2,\mathrm{min}}\right>$) have an input fidelity equal to $f$ (only that it is $\geq f$), we can easily construct a pair of pure states, $\left|\sigma'_{1,\mathrm{min}}\right>$ and $\left|\sigma'_{2,\mathrm{min}}\right>$, that achieve $F_{R,\mathrm{min}}(f)$ and have an input fidelity of $f$. Specifically, we can define
\begin{align}
    &\left|\sigma'_{1,\mathrm{min}}\right>=\left|\sigma_{1,\mathrm{min}}\right>\otimes \left|0\right>,\label{eq: min state 1}\\
    &\left|\sigma'_{2,\mathrm{min}}\right>=\left|\sigma_{2,\mathrm{min}}\right>\otimes \left(\frac{f}{f'}\left|0\right>+\sqrt{1-\left(\frac{f}{f'}\right)^2}\left|1\right>\right),\label{eq: min state 2}
\end{align}
where the final qubit simply gives a constant multiplicative factor of $\frac{f}{f'}$ to both the input and the output fidelity. These states also achieve the minimum possible output fidelity for a given input fidelity.

\section{Proof of the minimum relative fidelity continuity bound}\label{appendix continuity}

We start by finding a continuity bound for relative fidelity, before using it to find the continuity bounds for minimum relative fidelity. Given a pair of density matrices, $\{\sigma_1',\sigma_2'\}$, with a relative fidelity of $F_R(\sigma_1',\sigma_2')$, we wish to bound the relative fidelity of a nearby pair of density matrices, $\{\sigma_1,\sigma_2\}$. By nearby, we mean that the Bures distances, $d_B(\sigma_1,\sigma_1')$ and $d_B(\sigma_2,\sigma_2')$, between $\sigma_1$ and $\sigma_1'$ and between $\sigma_2$ and $\sigma_2'$ are known. We define
\begin{align}
    &\delta(\sigma_1,\sigma_1',\sigma_2,\sigma_2')=\frac{1}{\sqrt{2}}\left(d_B(\sigma_1,\sigma_1')+d_B(\sigma_2,\sigma_2')\right),\label{eq: delta def}\\
    &d_B(\rho_1,\rho_2)=\sqrt{2}\sqrt{1-F(\rho_1,\rho_2)}.
\end{align}

Starting from the triangle inequalities for the Bures distance, we can derive similar relationships for the fidelity. The triangle inequalities tell us that
\begin{align}
    &d_B(\sigma_1,\sigma_2)\leq d_B(\sigma_1',\sigma_2')+\sqrt{2}\delta(\sigma_1,\sigma_1',\sigma_2,\sigma_2'),\\
    &d_B(\sigma_1,\sigma_2)\geq d_B(\sigma_1',\sigma_2')-\sqrt{2}\delta(\sigma_1,\sigma_1',\sigma_2,\sigma_2').
\end{align}
We can therefore derive
\begin{align}
    &F(\sigma_1,\sigma_2)\geq 1 - \left(\sqrt{1-F(\sigma_1',\sigma_2')}+\delta\right)^2,\label{eq: fidelity triangle LB}\\
    &F(\sigma_1,\sigma_2)\leq 1 - \left(\sqrt{1-F(\sigma_1',\sigma_2')}-\delta\right)^2.\label{eq: fidelity triangle UB}
\end{align}
Note that Eqs.~(\ref{eq: fidelity triangle LB}) and (\ref{eq: fidelity triangle UB}) apply to any two pairs of states $\{\sigma_1,\sigma_2\}$ and $\{\sigma_1',\sigma_2'\}$. In order to lower bound the relative fidelity, $F_R(\sigma_1,\sigma_2)$, we must upper bound the input fidelity, $F(\sigma_1,\sigma_2)$, and lower bound the output fidelity, $F_{\mathrm{out}}(\sigma_1,\sigma_2)$.

The upper bound on the input fidelity comes directly from Eq.~(\ref{eq: fidelity triangle UB}). We now note that
\begin{align}
    &d_B(\sigma_1,\sigma_1')\geq d_B(\mathcal{I}\otimes\mathcal{C}_1[\sigma_1],\mathcal{I}\otimes\mathcal{C}_1[\sigma_1']),\label{eq: monot 1}\\
    &d_B(\sigma_2,\sigma_2')\geq d_B(\mathcal{I}\otimes\mathcal{C}_2[\sigma_2],\mathcal{I}\otimes\mathcal{C}_2[\sigma_2']),\label{eq: monot 2}
\end{align}
from the monotonicity of the Bures distance. Therefore, we can write the lower bound
\begin{equation}
    F_{\mathrm{out}}(\sigma_1,\sigma_2)\geq 1 - \left(\sqrt{1-F_{\mathrm{out}}(\sigma_1',\sigma_2')}+\delta\right)^2.\label{eq: out_fid continuity}
\end{equation}
This comes from substituting the output states for the input states in Eq.~(\ref{eq: fidelity triangle LB}) and applying the fact that the distance between the two pairs of output states cannot be greater than the distance between the two pairs of input states (due to Eqs.~(\ref{eq: monot 1}) and (\ref{eq: monot 2})). Consequently, we can write the continuity bound:
\begin{equation}
    F_R(\sigma_1,\sigma_2)\geq\frac{F_{\mathrm{out}}(\sigma_1',\sigma_2')-2\delta\sqrt{1-F_{\mathrm{out}}(\sigma_1',\sigma_2')}+\delta^2}{F(\sigma_1',\sigma_2')+2\delta\sqrt{1-F(\sigma_1',\sigma_2')}-\delta^2}.\label{eq: rel_fid continuity}
\end{equation}

This gives us a way of numerically finding $F_{R,\mathrm{min}}(f)$. We know that $F_{R,\mathrm{min}}(f)$ will be achieved by some pair of pure, $d^2$-dimensional qudit states that have a fidelity greater than or equal to $f$. Further, the set of pairs of density matrices that have a fidelity greater than or equal to $f$, $\{\mathcal{D}(d^2):F(\sigma_1,\sigma_2)\geq f\}$, is convex. In other words, given two pairs of states, $\{\sigma_1^{(A)},\sigma_2^{(A)}\}$ and $\{\sigma_1^{(B)},\sigma_2^{(B)}\}$, both of which lie in $\{\mathcal{D}(d^2):F(\sigma_1,\sigma_2)\geq f\}$, any convex combination of the pairs also lies in $\{\mathcal{D}(d^2):F(\sigma_1,\sigma_2)\geq f\}$. We can write
\begin{equation}
    F(p\sigma_1^{(A)}+(1-p)\sigma_1^{(B)},p\sigma_2^{(A)}+(1-p)\sigma_2^{(B)})\geq f,
\end{equation}
where we have used the concavity of the Bures fidelity. The set of pairs of pure states with fidelity greater than or equal to $f$ forms part of the boundary of this convex set and is therefore compact. Therefore, by taking a finite number of samples from the set of pairs of pure states with fidelity greater than or equal to $f$, bounding the Bures distance between the samples, and using Eq.~(\ref{eq: rel_fid continuity}) to lower bound the relative fidelity of any point between the samples, we can numerically lower bound the relative fidelity for any non-zero $f$ (for $f=0$, this method will not give a non-trivial bound on $F_{R,\mathrm{min}}(0)$; instead $F_{R,\mathrm{min}}(0)$ should be found analytically, as the limit of a sequence). By increasing the number of samples, we can tighten this lower bound, which will tend towards the true value of $F_{R,\mathrm{min}}(f)$ asymptotically with the number of samples. This concept is explored in more detail in Appendix~\ref{appendix algorithm}, where we present an algorithm for a numerical method to calculate $F_{R,\mathrm{min}}(f)$.

We now bound the behaviour of $F_{R,\mathrm{min}}(f)$ as a function of $f$. First, let us note that any pair of states, $\{\sigma_1,\sigma_2\}$, with a fidelity of $f$ has a distance from some different pair of states, $\{\sigma_1',\sigma_2'\}$, with a fidelity greater than or equal to $f'$ that is upper bounded by
\begin{equation}
    \delta(\sigma_1,\sigma_1',\sigma_2,\sigma_2')\leq \sqrt{|f'-f|}.\label{eq: fid dif to distance}
\end{equation}
To show this, we need only prove that there always exists such a pair, $\{\sigma_1',\sigma_2'\}$. Let us assume that $f<f'$ (since otherwise this is trivially true). Set
\begin{equation}
    \sigma_1'=\sigma_1
\end{equation}
and set
\begin{equation}
    \sigma_2'=p \sigma_1 + (1-p) \sigma_2.
\end{equation}
Then, recalling that fidelity is concave, we write
\begin{equation}
    \begin{split}
        F(\sigma_1',\sigma_2') &\geq p F(\sigma_1,\sigma_1) + (1-p) F(\sigma_1,\sigma_2)\\
        &\geq p + (1-p)f.
    \end{split}
\end{equation}
We now set
\begin{equation}
    p = \frac{f'-f}{1-f},
\end{equation}
which satisfies
\begin{equation}
    p + (1-p)f = f',
\end{equation}
as required. Next, we calculate
\begin{equation}
    \begin{split}
        F(\sigma_2,\sigma_2') &\geq p F(\sigma_1,\sigma_2) + (1-p) F(\sigma_2,\sigma_2)\\
        &\geq p f + (1-p)
        = 1 - (f'-f).
    \end{split}
\end{equation}
Finally, from Eq.~(\ref{eq: delta def}), we get
\begin{equation}
    \begin{split}
        \delta(\sigma_1,\sigma_1',\sigma_2,\sigma_2')&=\left(\sqrt{1-F(\sigma_1,\sigma_1')}+\sqrt{1-F(\sigma_2,\sigma_2')}\right)\\
        &\leq \sqrt{f'-f},
    \end{split}
\end{equation}
completing the proof.

Let us stipulate that $\{\sigma_1,\sigma_2\}$ are states that achieve a relative fidelity of $F_{R,\mathrm{min}}(f)$. Combining Eqs.~(\ref{eq: out_fid continuity}) and (\ref{eq: fid dif to distance}), we have
\begin{equation}
    F_{\mathrm{out}}(\sigma_1,\sigma_2)\geq 1 - \left(\sqrt{1-F_{\mathrm{out}}(\sigma_1',\sigma_2')}+\sqrt{f'-f}\right)^2.
\end{equation}
Then, noting that
\begin{equation}
    F_{\mathrm{out}}(\sigma_1',\sigma_2')\geq f' F_{R,\mathrm{min}}(f'),
\end{equation}
we can write the continuity bound in Eq.~(\ref{eq: min_rel_fid continuity}), which we repeat here for clarity:
\begin{equation*}
    F_{R,\mathrm{min}}(f)\geq \frac{1 - \left(\sqrt{1-f' F_{R,\mathrm{min}}(f')} +\sqrt{f'-f}\right)^2}{f}.
\end{equation*}

\section{Algorithm for numerically finding the minimum relative fidelity}\label{appendix algorithm}

\begin{figure}[t]
\centering
\includegraphics[width=0.8\linewidth]{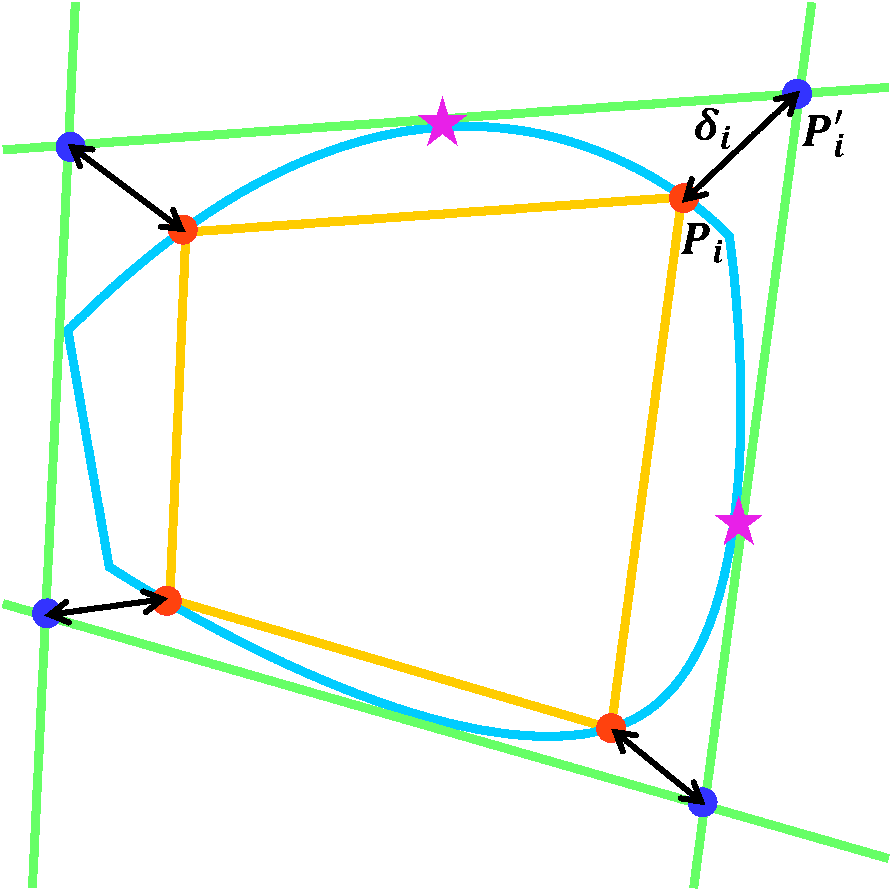}\caption{Illustration of the algorithm for numerically finding the minimum relative fidelity. This diagram shows how the process would work in two dimensions (although, even in the qubit channel case, the actual dimension of the set we are minimising over is much more than two). The pale blue outline is the surface of the set, $\mathcal{S}$, of valid points (pairs of valid density matrices with a pairwise fidelity greater than $f$). We make no assumptions about the geometry of this set, other than that it is convex (meaning that the method works even if the surface of the set is not smooth, as in this diagram). The red dots on the surface of $\mathcal{S}$ are points, $\{P_i\}$, for which we know the value of the output fidelity. The planes (lines) joining each known point to its neighbours define a polygon $\mathcal{R}$ within the set $\mathcal{S}$, the surface of which is given by the orange lines. The green lines that are tangential to the set $\mathcal{S}$ are also parallel to the orange lines and intersect each other at the points $\{P'_i\}$, which are represented by dark blue dots. Finally, we define the polygon $\mathcal{Q}$ as the region enclosed by the green lines. $\mathcal{Q}$ contains $\mathcal{S}$, meaning there are no points in $\mathcal{S}$ that are not also in $\mathcal{Q}$. Consequently, no point in $\mathcal{S}$ is more distant from $\mathcal{R}$ (in terms of the trace norm between the density matrices comprising the two points) than the most distant point in $\mathcal{Q}$. In fact, the points $\{P'_i\}$ are the most distant points from $\mathcal{R}$ in $\mathcal{Q}$. For each known point $P_i$, we must deduct an error cost, based on the distance $\delta_i$, from the value of the output fidelity at that point, in order to lower bound the minimum output fidelity. We can now detail the update rule. Suppose the point labelled $P_i$ on the diagram is the one that gives rise to the lower bound on the minimum output fidelity. Then, the candidate points (represented by the pink stars) are points at which the green lines that pass through point $P'_i$ touch $\mathcal{S}$. Whichever of these is most distant from $P_i$ is picked and added to the set of known points. We recalculate the points $\{P'_i\}$ and the distances $\{\delta_i\}$ accordingly.}
\label{fig: concave minimisation}
\end{figure}

Due to the continuity relation given in Eq.~(\ref{eq: rel_fid continuity}), we can numerically bound the minimum relative fidelity (for a given, finite input fidelity, $f$) from below by sampling sufficiently many points. One way of doing this would be to parametrise the set of pairs of pure states with an input fidelity greater than or equal to $f$ and then to evenly sample the parameters, calculating both the relative fidelity for that set of parameters and the maximum distance between any pair of pure states (with input fidelity greater than or equal to $f$) and the nearest sampled state. This would, however, be very inefficient. Instead, we present an algorithmic method that converges to the true value of $F_{R,\mathrm{min}}(f)$ as the number of samples increases.

We begin by recalling that the output fidelity is a concave function that can be minimised by a pair of states of the form given in Eqs.~(\ref{eq: min state 1}) and (\ref{eq: min state 2}). This means that the minimum output fidelity is always achievable by a minimisation over all pairs of $2d^2$-dimensional qudit states. In fact, we can constrain the final qubit of the first state to be $\left|0\right>\left<0\right|$, as per Eqs.~(\ref{eq: min state 1}) and (\ref{eq: min state 2}), resulting in a minimisation over a $d^2$-dimensional qudit state and a $2d^2$-dimensional qudit state and reducing the dimension of the problem by $3d^4$; for simplicity, we do not do this here, but the algorithm can be trivially changed accordingly. Expanding Eq.~(\ref{eq: out_fid continuity}) in terms of $\delta$, we get
\begin{equation}
    \left|F_{\mathrm{out}}(\sigma_1',\sigma_2') - F_{\mathrm{out}}(\sigma_1,\sigma_2)\right|\leq 2\delta-\delta^2.\label{eq: out_fid diff}
\end{equation}
Note that this continuity relation, whilst not tight, is written purely in terms of $\delta$.

We will refer to a pair of Hermitian, $2d^2$-dimensional, unit trace, square matrices as a point. Consider the set of pairs of pure states with an input fidelity equal to $f$. Let us call the convex hull of this set (the set of all convex combinations of such points) $\mathcal{S}$. $\mathcal{S}$ is the convex set of points that we must minimise the output fidelity (a concave function) over. The dimension of $\mathcal{S}$ is $D=2(4d^4-1)$, in that this is the minimum number of real coordinates required to map each pair of density matrices to a unique displacement from the origin in coordinate space. If $P$ is the matrix pair $\{\sigma_1,\sigma_2\}$, we define
\begin{equation}
    F_{\mathrm{out}}(P) = F_{\mathrm{out}}(\sigma_1,\sigma_2).
\end{equation}

We can define a hyperplane in $\mathcal{S}$ as the set of points, $P$, that can be written as
\begin{equation}
    P = P_0 + \sum_{i=1}^{D-1} p_i V_i,\label{eq: plane eq}
\end{equation}
where $P_0$ is a point and the $V_i$ are pairs of Hermitian, $2d^2$-dimensional, trace zero, square matrices, which play the role of vectors. Summation, in this context, means adding the first matrices of each pair to each other and then doing the same for the second matrices of each pair. The set $\{V_i\}$ therefore defines the hyperplane. The plane can equivalently be defined as the set of all points for which $\mathrm{Tr}[N P]$ has a constant value, where $N$ is a pair of Hermitian, $2d^2$-dimensional, trace zero, square matrices, which we call the normal to the hyperplane. We can find $N$ from $\{V_i\}$ by solving the $D-1$ simultaneous equations
\begin{equation}
    \mathrm{Tr}[N V_i]=0~~\forall i \in (1,D-1).
\end{equation}
Any point on the boundary of $\mathcal{S}$, $P_{\mathrm{bound}}$, can be written as
\begin{equation}
    P_{\mathrm{bound}} = \argmax_{P\in \mathcal{S}} \mathrm{Tr}[N P],\label{eq: supporting hyperplane}
\end{equation}
for some choice of $N$ (this is a version of the supporting hyperplane theorem, and is basically a statement that the boundary of a convex set is the furthest you can go in a given direction whilst remaining within the set). Equally, any point expressible as in Eq.~(\ref{eq: supporting hyperplane}) is on the boundary. The plane defined by this $N$ is a tangent to the set $\mathcal{S}$.

Now note that a set of points, $\mathcal{P}=\{P_i\}$, define a polygon, the interior of which is comprised of all points expressible as
\begin{equation}
    P = \sum_{i} p_i P_i,
\end{equation}
where the $\{p_i\}$ define a convex combination (i.e. they are all non-negative and sum to $1$) and the index $i$ ranges over the labels of all of the points in $\mathcal{P}$. The surface of this polygon is made up of hyperplanes and, so long as all of the points are linearly independent (no point can be written as a convex combination of the others), each point on a given face can be written as a convex combination of only $D-1$ of the points in $\mathcal{P}$. Within a polygon, a concave function is minimised at one of the vertices, so the minimum value of $F_{\mathrm{out}}$ over the polygon will be equal to
\begin{equation*}
    \min_{{P\in \mathcal{P}}}F_{\mathrm{out}}(P).
\end{equation*}

We can now briefly outline an algorithm to numerically find $F_{\mathrm{out},\mathrm{min}}(f)$ (and hence $F_{R,\mathrm{min}}(f)$). The basic idea is to find the minimum output fidelity for a finite number of points on the boundary of $\mathcal{S}$. By doing so, we find the minimum over a (polygonal) subset of $\mathcal{S}$, which we call $\mathcal{R}$. We then upper bound the distance ($\delta$ in Eq.~(\ref{eq: out_fid diff})) between any point in $\mathcal{S}$ and the nearest point in $\mathcal{R}$. We do this by finding a different convex, polygonal set, $\mathcal{Q}$, which surrounds $\mathcal{S}$ (in the sense that every point in $\mathcal{S}$ is in $\mathcal{Q}$) and finding the distance from each vertex of $\mathcal{R}$ to the corresponding vertex of $\mathcal{Q}$. Finally, we use Eq.~(\ref{eq: out_fid diff}) to lower bound the minimum output fidelity for any point in $\mathcal{S}$. Since $\mathcal{S}$ includes all pairs of pure states of dimension $2d^2$, we can therefore also lower bound the minimum relative fidelity by dividing the minimum output fidelity by $f$. Fig.~\ref{fig: concave minimisation} provides a visualisation of how the algorithm works (albeit for only two dimensions).

\begin{enumerate}
    \item Pick $D+1$ initial points (pairs of states) from the set of pairs of states with an input fidelity equal to $f$. The initial set of points (which we will call $\mathcal{P}=\{P_i\}$) can be chosen in any way, so long as the points are linearly independent. One way of doing this would be to parametrise pure qudit states by a set of angles, which could then be randomly chosen. All of the points in $\mathcal{P}$ will lie on the boundary of $\mathcal{S}$. $\mathcal{P}$ defines the polygon $\mathcal{R}$ (each point is a vertex), the interior of which is comprised of all convex combinations of the chosen points.
    
    \item Calculate the output fidelity for each point in $\mathcal{P}$. Write $F_{\mathrm{out},i}=F_{\mathrm{out}}(P_i)$. From the concavity of the output fidelity, the minimum output fidelity for any point in $\mathcal{R}$ is given by $\min_i F_{\mathrm{out},i}$.
    
    \item Let $\mathcal{F}$ be the set of faces of $\mathcal{P}$ (that is, the set of hyperplanes that make up the surface of $\mathcal{R}$). Initially, there will be $D+1$ faces (the same as the number of points in $\mathcal{P}$). Each $F_i$ can be expressed as in Eq.~(\ref{eq: plane eq}) by setting $P_0=P_j$ and the $V$ ``vectors" to $P_j-P_k$, where the $j$ and $k$ refer to some arbitrary labelling of the points that form the vertices of $F_i$. For each face, $F_i$, in $\mathcal{F}$, find the normal to it, $N_i$. Some degree of care is required in choosing the sign of $N_i$ so that it points outwards from the interior of $\mathcal{R}$. Then, for each $N_i$, find
    \begin{align}
        &\mu_i= \max_{P\in\mathcal{S}}\mathrm{Tr}[N_i P],\\
        &P_{\mathrm{max},i}= \argmax_{P\in\mathcal{S}}\mathrm{Tr}[N_i P]
    \end{align}
    The points $P_{\mathrm{max},i}$ will be on the boundary, as per Eq.~(\ref{eq: supporting hyperplane}). This is a maximisation of a linear function over a convex set and can be done by, for instance, semidefinite programming.
    
    \item We now define a new set of hyperplanes, $\mathcal{F}'$, comprised of hyperplanes, $F'_i$, that are parallel to the faces $F_i$ but that pass through the points $P_{\mathrm{max},i}$. These hyperplanes are therefore tangential to the set $\mathcal{S}$. We can then define the set, $\mathcal{P}'$, of points at which each hyperplane intersects with each of its $D-1$ nearest neighbours. These points then define a new polygon, $\mathcal{Q}$, that surrounds the set $\mathcal{S}$ (since each of its faces are tangential to the set). No point in $\mathcal{S}$ can be more distant from a point in $\mathcal{F}$ than the vertices of $\mathcal{Q}$ (i.e. the points in  $\mathcal{P}'$). Thus, by finding the distance between each point $P_i$ and the corresponding point $P'_i$, we can upper bound $\delta$ at each point. One complication is that the Bures distance is not a valid distance metric outside of $\mathcal{S}$. However, we can get around this by using the trace norm (which is a valid distance metric for all Hermitian matrices) and then bounding the Bures distance using the Fuchs van der Graaf relations. In other words, by finding the trace norm between $P_i$ and $P'_i$, we bound the maximum trace norm from $\mathcal{R}$ for any point lying between $\mathcal{R}$ and $\mathcal{Q}$, which includes the entire boundary of $\mathcal{S}$, and then this upper bound on the trace norm is converted into an upper bound on the Bures distance.
    
    \item For each pair $\{A,B\}$ in $\mathcal{P}$ and corresponding pair $\{A',B'\}$ in $\mathcal{P}'$, calculate
    \begin{equation}
        \delta_i=\sqrt{||A-A'||}+\sqrt{||B-B'||}.
    \end{equation}
    Next, calculate the cost function,
    \begin{equation}
        c_i=2\delta_i-\delta_i^2,
    \end{equation}
    where we have used the continuity relation from Eq.~(\ref{eq: out_fid diff}). Finally, for each point in $\mathcal{P}$, deduct $c_i$ from $F_{\mathrm{out},i}$. The smallest value of $c_i-F_{\mathrm{out},i}$ gives a lower bound on $F_{\mathrm{out},\mathrm{min}}(f)$, whilst the smallest value of $F_{\mathrm{out},i}$ gives an upper bound on $F_{\mathrm{out},\mathrm{min}}(f)$.
    
    \item We now detail the update rule for finding new bounds. The point giving rise to the current lower bound on $F_{\mathrm{out},\mathrm{min}}(f)$, which we will label $P_1$ for ease, touches $D$ faces. For each of these faces, we know the point at which a hyperplane parallel to them is tangential to $\mathcal{S}$. Pick the one that is furthest from $P_1$ and add it to $\mathcal{P}$. This will remove a face from $\mathcal{F}$ and add $D$ new ones. The vertices of each of the new faces are the vertices of the old face with one of them replaced by the new point each time. Repeat steps 3 to 5 for each of the new faces. 
    
    \item Repeat step 6 until the desired level of precision is achieved. Divide by the input fidelity, $f$, to obtain bounds on $F_{R,\mathrm{min}}(f)$.
\end{enumerate}

Since each iteration can only improve the lower bound on $F_{\mathrm{out},\mathrm{min}}(f)$ (or leave it unchanged) and $\mathcal{R}$ will model $\mathcal{S}$ increasingly well as the number of points increases, it is evident that the bound obtained using this algorithm will converge to the true value asymptotically in the number of iterations. We have no guarantee, however, about the rate of convergence.

\section{Minimum relative fidelity for specific channel pair examples}\label{appendix examples}

First, we consider discrimination between the identity channel and a Pauli channel. Specifically, we consider the channel that applies the Pauli-X operator with probability $p$ and applies the identity operator with probability $1-p$.

Consider the input states $\left|\phi_1\right>$ and $\left|\phi_2\right>$. The output fidelity is
\begin{equation}
    F_{\mathrm{out}}^{X}=\sqrt{(1-p)\left|\left<\phi_1\middle|\phi_2\right>\right|^2+p\left|\left<\phi_1\right|X\left|\phi_2\right>\right|^2},
\end{equation}
where $X$ is the Pauli-X operator. Since the input fidelity is $\left|\left<\phi_1\middle|\phi_2\right>\right|$, we can write
\begin{equation}
    F_{R}^{X}=\sqrt{(1-p)+p\frac{\left|\left<\phi_1\right|X\left|\phi_2\right>\right|^2}{\left|\left<\phi_1\middle|\phi_2\right>\right|^2}}.
\end{equation}
Setting both $\left|\phi_1\right>$ and $\left|\phi_2\right>$ to $\left|0\right>$ gives $\left|\left<\phi_1\right|X\left|\phi_2\right>\right|=0$ and so the minimum relative fidelity is given by
\begin{equation}
    F_{R,\mathrm{min}}^{X}=\sqrt{1-p},\label{eq: pauli FR}
\end{equation}
which is always independent of the input fidelity. This tells us that discrimination protocols for this pair of Pauli channels cannot benefit from adaptivity. This is as expected because all Pauli channels are jointly teleportation covariant~\cite{pirandola_fundamental_2017}.

Next, we apply our method to discrimination between unitary channels. In this case, we expect that the minimum relative fidelity will go to zero for some non-zero input fidelity, because unitary channels are perfectly discriminable after finite channel uses~\cite{acin_statistical_2001,dariano_using_2001}. We consider unitaries of the form
\begin{equation}
    U(\theta)=X^{\theta},\label{eq: U def}
\end{equation}
where $\theta$ is a real parameter. Since we are considering a pair of unitaries, $U(\theta_1)$ and $U(\theta_2)$, the quantity of interest is $\theta_1-\theta_2$, since the relative fidelity for a pair of channels is unchanged by a unitary applied before the channels (so any pair $\{\theta_1,\theta_2\}$ with the same value of $\theta_1-\theta_2$ will have the same minimum relative fidelity for any given input fidelity). Thus, for simplicity, we consider unitaries $U(0)$ (the identity channel) and $U(\theta)$.

The eigenvalues of a unitary, $U$, all have magnitude $1$, and so can be expressed as $\exp(i\phi^U_j)$, where $j$ is a label for the eigenvalue that runs from $1$ to the dimension of the unitary, $d$. Suppose the eigenvalues are ordered such that $\phi^U_j\leq\phi^U_{j+1}$ (with all of the $\phi_j$ confined to the range $[0,2\pi)$). For ease, define $\phi^U_{\mathrm{min}}=\phi^U_{1}$ and $\phi^U_{\mathrm{max}}=\phi^U_{d}$. From Ref.~\cite{yuan2017quantum}, we know that $F_{\mathrm{con}}$ for a unitary and the identity channel is given by
\begin{equation}
    F_{\mathrm{con}}^{U}=\cos\left(\frac{\phi^U_{\mathrm{max}}-\phi^U_{\mathrm{min}})}{2}\right),\label{eq: fcon U}
\end{equation}
so long as $F_{\mathrm{con}}^{U}\geq 0$.

Note that the minimum relative fidelity can be expressed as
\begin{equation}
    F_{R,\mathrm{min}}^{U}=\frac{1}{f}\min_{\left|\psi\right>}\min_{u}|\left<\psi\middle|u (\mathcal{I}\otimes U)\middle|\psi\right>|,
\end{equation}
where the identity acts on any idler modes and $u$ is a different unitary that obeys the constraint
\begin{equation}
    |\left<\psi\middle|u\middle|\psi\right>| \geq f.
\end{equation}
This is because we can define the input state for the unitary as $\left|\psi\right>$ and then express the input state for the identity channel as $u^{\dagger}\left|\psi\right>$. Using Eq.~(\ref{eq: fcon U}), we can write
\begin{equation}
    \cos\left(\frac{\phi^u_{\mathrm{max}}-\phi^u_{\mathrm{min}})}{2}\right)\geq f.
\end{equation}

The eigenvalues of the product of two unitaries, $u(\mathcal{I}\otimes U)$, are constrained by~\cite{childs_quantum_2000,chau_elementary_2011}
\begin{equation}
    \phi^{u (\mathcal{I}\otimes U)}_{\mathrm{min}} \geq \phi^{u}_{\mathrm{min}} + \phi^{U}_{\mathrm{min}},~~\phi^{u(\mathcal{I}\otimes U)}_{\mathrm{max}} \leq \phi^{u}_{\mathrm{max}} + \phi^{U}_{\mathrm{max}}.\label{eq: eigenvalue inequalities}
\end{equation}
For $U(\theta)$, as defined in Eq.~(\ref{eq: U def}), we have
\begin{equation}
    \phi^{U}_{\mathrm{min}} = 0,~~\phi^{U}_{\mathrm{max}} = \pi\theta.
\end{equation}
Hence, we can write
\begin{equation}
    F_{R,\mathrm{min}}^{U}\geq \frac{1}{f}\cos\left(\frac{\pi\theta}{2}+\arccos(f)\right),
\end{equation}
where $\arccos(f)$ takes its principle value. If we let $u$ also be of the form in Eq.~(\ref{eq: U def}), the inequalities in Eq.~(\ref{eq: eigenvalue inequalities}) become equalities, so we can write
\begin{equation}
    F_{R,\mathrm{min}}^{U} = \frac{1}{f}\cos\left(\frac{\pi\theta}{2}+\arccos(f)\right),\label{eq: unitary FR}
\end{equation}
where we assume the right-hand side is non-negative (if not, we set it to $0$). This corresponds to an output fidelity bound, after $N$ channel uses, of
\begin{equation}
    F_{N}^{U}\geq\cos\left(\frac{N\pi\theta}{2}\right)
\end{equation}
for $N<\theta^{-1}$ (and zero for $N>\theta^{-1}$).

Finally, we consider a pair of entanglement-breaking, qubit channels, $\mathcal{C}_1^{\mathrm{EB}}$ and $\mathcal{C}_2^{\mathrm{EB}}$, defined by a measurement along one of a pair of axes followed by a rotation. Specifically, $\mathcal{C}_i^{\mathrm{EB}}$ consists of the positive operator-valued measurement
\begin{align*}
    &M(\theta_i)=\{\left|\phi(\theta_i)\right>\left<\phi(\theta_i)\right|,\mathcal{I}-\left|\phi(\theta_i)\right>\left<\phi(\theta_i)\right|\},\\
    &\left|\phi(\theta)\right>=\cos(\theta)\left|0\right>+\sin(\theta)\left|1\right>,
\end{align*}
followed by the rotation given by
\begin{equation*}
    R(\theta_i)=\begin{pmatrix}
    \cos(\theta_i) &\sin(\theta_i)\\
    -\sin(\theta_i) &\cos(\theta_i)
    \end{pmatrix}.
\end{equation*}
Once again, it is only the difference in rotation angle, $\Delta_\theta=\theta_1-\theta_2$, that matters, since we will get the same minimum relative fidelity for any value of $\theta_1$ (so long as $\Delta_\theta$ is fixed). We can therefore set $\theta_1=0$ without loss of generality.

This pair of channels is of interest because we expect adaptivity to be of benefit (this can be numerically confirmed; see the supplementary Mathematica files~\cite{SUPP}), but we also expect that the minimum relative fidelity will never go to zero (except in the case where $|\Delta_\theta|=\frac{n}{4}\pi$, for integer $n$), because no non-orthogonal input states will result in orthogonal output states. This is in line with the fact that discrimination strategies between classical channels can benefit from adaptivity but adaptivity cannot improve the asymptotic rate of decay of the discrimination error probability~\cite{hayashi_discrimination_2009}. It is also known that if two classical channels are not perfectly discriminable after a single channel use, they will not be perfectly discriminable after any finite number of channel uses~\cite{harrow_adaptive_2010}.

Since the channels are entanglement-breaking, we only need to consider single qubit inputs (without idlers), because idlers cannot decrease the relative fidelity in this case. This reduces the difficulty of the minimisation.

The minimum relative fidelity if the same state is used for both channels is given by
\begin{equation}
    F_{\mathrm{con}}^{\mathrm{EB}}=\frac{1}{2}\sqrt{2+\cos(2\Delta_\theta)+\cos(6\Delta_\theta)}.
\end{equation}
As expected, $F_{\mathrm{con}}^{\mathrm{EB}}\to 1$ as $\Delta_\theta\to 0$. Minimising the relative fidelity over all pairs of single qubit states (with no constraint on the input fidelity), we find, under the numerically verifiable assumption that we can set one of the input states to $\left|0\right>$, that the minimum relative fidelity is given by
\begin{equation}
    F_{R,\mathrm{min}}^{\mathrm{EB}}(0)=\frac{|\cos(\Delta_\theta)+\cos(3\Delta_\theta)|}{\sqrt{2+2\cos(2\Delta_\theta)+\cos(4\Delta_\theta)}}.\label{eq: classical FR}
\end{equation}
This is less than $F_{\mathrm{con}}^{\mathrm{EB}}$ (except for at certain values of $\Delta_\theta$), meaning that discrimination protocols can benefit from adaptivity, but only goes to zero for six different values of $\Delta_\theta$ (in the range $0\leq\Delta_\theta<2\pi$). It is interesting to note that $F_{R,\mathrm{min}}^{\mathrm{EB}}(0)\not\to 1$ as $\Delta_\theta\to 0$. In fact, $F_{R,\mathrm{min}}^{\mathrm{EB}}(0)\to \frac{2}{\sqrt{5}}$. This may initially seem strange, as the channels are identical for $\Delta_\theta=0$, however, as the separation between the channels, $\Delta_\theta$, becomes infinitesimal, the input fidelity required in order to achieve the minimum relative fidelity becomes infinitesimal too. The input fidelity between the optimal states achieving the minimum relative fidelity is given by
\begin{equation}
    f^{\mathrm{EB}}_{\mathrm{opt}}=\frac{1-\cos(2\Delta_\theta)\cos(4\Delta_\theta)}{|\sin
    (\Delta_\theta)|\sqrt{4+2\cos(2\Delta_\theta)-2\cos(6\Delta_\theta)}},\label{eq: classical required in fid}
\end{equation}
which approaches zero as $\Delta_\theta\to 0$. Thus, as $\Delta_\theta$ becomes small, the number of previous channel uses required in order to have sufficiently separated input states to achieve the minimum possible relative fidelity, $F_{R,\mathrm{min}}^{\mathrm{EB}}(0)$, with the next channel use becomes large. Another feature of interest is that $F_{R,\mathrm{min}}^{\mathrm{EB}}(0)=F_{\mathrm{con}}^{\mathrm{EB}}\neq 0$ for $\Delta_\theta=\frac{(1+2x)\pi}{8}$, where $x$ is an integer. This shows that adaptivity does not have any benefit for these parameter values.

We can bound the output fidelity of any adaptive protocol as
\begin{equation}
    F^{\mathrm{EB}}_N\geq F_{R,\mathrm{min}}^{\mathrm{EB}}(0)^N,
\end{equation}
which is equivalent to the bound obtained using the amortised channel divergence from Ref.~\cite{wilde_amortized_2020}, however this bound can be much looser (for small $\Delta_{\theta}$) than one obtained by calculating the relative fidelity for the $N$-th channel use recursively (i.e. using Eq.~(\ref{eq: adaptive LB})).

To generate Fig.~\ref{fig: scaling}, Eqs.~(\ref{eq: pauli FR}), (\ref{eq: unitary FR}), and (\ref{eq: classical FR}) were used, with the channel parameters ($p$, $\theta$, and $\Delta_\theta$) chosen such that $F_{\mathrm{con}}=0.95$ for all three.

See the supplementary Mathematica files for further details~\cite{SUPP}.

\section{Relation to previous works}\label{appendix differences}
Previous research into the benefit of adaptivity has used quantities similar to the minimum relative fidelity. Ref.~\cite{chiribella_quantum_2019} introduced the fidelity divergence, which is defined in the same way as the minimum relative fidelity but without the constraint on the input fidelity (i.e. it is equal to $F_{R,\mathrm{min}}(0)$). It was applied to the problem of identifying cause-effect relations and the fact that the input states can be restricted to the pure states was proved. Ref.~\cite{wilde_amortized_2020} introduced the amortised channel divergence, which is defined for a generalised divergence (i.e. it is defined for a whole class of functions that obey a data processing inequality). If this divergence is chosen to be the sandwiched R\'{e}nyi entropy with $\alpha=\frac{1}{2}$, then the resulting quantity is $-\ln[{F_{R,\mathrm{min}}(0)}]$. The asymptotic error exponent was given in terms of this quantity.

The key difference between the $F_{R,\mathrm{min}}(f)$, as defined in this work, and both of the previous quantities is the input fidelity constraint. This is what allows us to see how the discriminative power of a protocol increases with the number of uses, as the set of possible pairs of input states becomes larger. The existence of some initial distance between the states sent into each channel can allow the channels to produce output states with an even greater distance between them, however we must first generate that initial difference between the states with preceding channel uses. At the beginning of the protocol (i.e. for the first few channel uses), the states will not be far enough apart to obtain the full benefit that adaptivity (or entanglement between probe states) will later give for a larger number of channel uses. This means that asymptotic bounds that do not account for these initial channel uses will be looser than those that do.

As an example, consider the entanglement-breaking channels example, discussed in greater detail in Appendix~\ref{appendix examples}. Eq.~(\ref{eq: classical FR}) gives the minimum relative fidelity (without an input fidelity constraint) for this channel pair, whilst Eq.~(\ref{eq: classical required in fid}) gives the maximum input fidelity for a pair of input states that achieve this relative fidelity. If $\Delta_{\theta}$ is small, the number of channel uses required before we can have a pair of input states that are sufficiently far apart to achieve the minimum relative fidelity in Eq.~(\ref{eq: classical FR}) can be very large. The output fidelity bound obtained by nesting $F_{R,\mathrm{min}}(f)$, starting from $f=1$, until we have an output fidelity lower than the right-hand side of Eq.~(\ref{eq: classical required in fid}) and only then multiplying by $F_{R,\mathrm{min}}(0)$ for the remaining channel uses can be much smaller than the bound obtained by using $F_{R,\mathrm{min}}(0)^N$. As $\Delta_{\theta}$ becomes smaller, the difference becomes more pronounced. See the supplementary Mathematica file for more details~\cite{SUPP}.

It is the scaling of $F_{R,\mathrm{min}}(f)$ with $f$ (specifically, the calculation of the maximum possible scaling with $f$) that allows us to confirm that the maximum QFI is $N^2$ times the maximum one-shot QFI. Further, if $F_{R,\mathrm{min}}(0)=0$, we cannot give any useful channel bounds using only $F_{R,\mathrm{min}}(0)$. Using $F_{R,\mathrm{min}}(f)$, however, we can calculate the minimum number of channel uses that must occur before the channels are perfectly distinguishable and can give (non-zero) bounds on the output fidelity for protocols with fewer than this number of uses.

Finally, we show an important property of $F_{R,\mathrm{min}}(f)$, which holds in the case that $f=0$, that was not shown in previous works. It was noted in Ref.~\cite{wilde_amortized_2020} that the dimension of the states used to calculate $F_{R,\mathrm{min}}(0)$ (there called the fidelity divergence), or the equivalent amortised channel divergence, is unbounded; here we show that we can restrict the optimisation to be over pure states of dimension $d^2$. This makes finding the minimum a far more tractable problem than minimising over a pair of states with no restriction on the size of the idler system.

\end{document}